\documentclass[runningheads]{llncs}
\usepackage[T1]{fontenc}
\usepackage{comment}
\usepackage{algorithm}
\usepackage{algpseudocode}
\usepackage{amsmath}
\usepackage{soul}
\usepackage{url}
\usepackage[utf8]{inputenc}
\usepackage{graphicx}
\usepackage{booktabs}
\usepackage{subcaption}
\usepackage{enumitem}
\usepackage{xcolor}
\usepackage{multirow}
\usepackage{url}

\def\L{{\cal L}} 

\def\AE{{\sc ae}}
\def\L{{\sc \`{L}}}
\def\E{{\sc \`E}}
\def\D{{\sc \`D}}

\def\Val{{\sc v}}
 \def\Aro{{\sc a}}
\def\Dom{{\sc d}}

\newcommand{\edrl}{\textit{EDRL}} 
\newcommand{\mea}{\textit{MEA}}

\def\train{\kappa}

\def\Feature{{Biomarker}}

\def\RF{{\sc rf}}
\def\SVM{{\sc svm}}
\def\KNN{{\sc knn}}
\def\ANN{{\sc ann}}
\def\high{{\sc +}}
\def\low{{\sc -}}
\def\N#1{{Noise\_#1}}
\begin{document}
\title{Emotion-Disentangled Embedding Alignment for Noise-Robust and Cross-Corpus Speech Emotion Recognition}
%
\titlerunning{Noise-Robust Cross-Corpus SER via EDRL-MEA}
%
\author{Upasana Tiwari, Rupayan Chakraborty\orcidID{0000-0002-3566-0784}, \\
Sunil Kumar Kopparapu\orcidID{0000-0002-0502-527X}}
\authorrunning{Upasana Tiwari et al.}
%
\institute{TCS Research, Tata Consultancy Services Limited, INDIA 
\email{\{tiwari.upasana1,rupayan.chakraborty,sunilkumar.kopparapu\}@tcs.com}\\
}
\maketitle              
\begin{abstract}
Effectiveness of speech emotion recognition in real-world scenarios is often hindered by noisy environments and variability across datasets. This paper introduces a two-step approach to enhance the robustness and generalization of speech emotion recognition models through improved representation learning. First, our model employs \edrl\ (Emotion-Disentangled Representation Learning) to extract class-specific discriminative features while preserving shared similarities across emotion categories. Next, \mea\ (Multiblock Embedding Alignment) refines these representations by projecting them into a joint discriminative latent subspace that maximizes covariance with the original speech input. The learned \edrl-\mea\ embeddings are subsequently used to train an emotion classifier using clean samples from publicly available datasets, and are evaluated on unseen noisy and cross-corpus speech samples. Improved performance under these challenging conditions demonstrates the effectiveness of the proposed method.

\end{abstract}
\section{Introduction}

Speech Emotion Recognition (SER) is a vital area of research aimed at inferring the emotional state of a speaker, enabling machines to understand and respond to human emotions from speech signals. Accurate emotion detection supports the development of empathetic virtual assistants \cite{Mamun_2023}, responsive customer service agents \cite{abhishek2023wecareimprovingcode}, and other AI systems that interact naturally and contextually \cite{lin2024selfcontextawareemotionperception}. 

Despite recent progress in SER \cite{Schuller-STOA2018}, models often struggle with robustness and generalization, especially when exposed to \textit{unseen noise} and \textit{cross-corpus conditions} at inference time. These scenarios, where the model is tested on speech samples differing significantly from the clean, in-domain training data, reveal critical limitations in existing systems. Traditional approaches typically rely on fixed representations and static features that do not adapt well to real-world variations, including environmental distortions and dataset shifts.

A major obstacle lies in the variability of emotional expression across different datasets, shaped by cultural, linguistic, and speaker-specific differences \cite{rathi2024analyzing,ahn2021cross,parry2019analysis,braunschweiler2021study}. In parallel, background noise, such as babble, ambient disturbances, or acoustic corruption further degrades speech quality, making reliable feature extraction increasingly difficult \cite{george2024review,fahad2021survey,leem2022not,leem2023selective}. Existing SER methods often fail to generalize across such challenging acoustic and data conditions.

To address these limitations, we propose a two-step approach for robust and generalizable representation learning. Motivated from \cite{TiwariCK24}, first, we introduce an \textit{Emotion-Disentangled Representation Learning (\edrl)} framework that extracts class-specific discriminative features while retaining emotion-shared structures across categories. This disentangled representation promotes expressiveness while preserving generalizability, even when emotional cues vary across content and speakers. 

Second, we employ \textit{Multiblock Embedding Alignment (\mea)} to project the EDRL-derived embeddings into a joint latent space that aligns closely with the original speech input. \mea\ enhances the discriminative capacity of these features by maximizing shared covariance across blocks, allowing the model to better distinguish emotional states even under noisy or cross-corpus conditions.

The learned \edrl-\mea\ embeddings are then used to train an emotion classifier using clean samples from the IEMOCAP dataset. Evaluation is conducted on \textit{unseen noisy} and \textit{cross-corpus} test samples, demonstrating marked improvements in robustness and generalization over conventional methods. By jointly learning emotion-specific representations and refining them through projection alignment, our approach improves SER performance in diverse and unpredictable acoustic environments.

The key contributions of this paper are:
\begin{itemize}
    \item We make use of a two-stage SER framework based on \edrl-\mea\ that creates robust emotion embeddings effective in both clean and unseen noisy, cross-corpus scenarios.
    \item The \edrl-\mea\ architecture acts as a pre-trained embedding generator without requiring any fine-tuning, domain adaptation, or data augmentation, enabling simplicity alongside improved generalization.
    \item Our method effectively captures emotion-specific discriminative patterns and refines them through embedding alignment, making it suitable for real-world, variable conditions.
\end{itemize}

The remainder of the paper is organized as follows: Section~\ref{sec:review} reviews related work. Section~\ref{sec:method} details our proposed \edrl-\mea\ methodology. Section~\ref{sec:exp} presents experiments and analysis. Section~\ref{sec:conclu} concludes the paper.

\section{Literature Review}
\label{sec:review}

Recent advances in deep learning have significantly influenced the field of Speech Emotion Recognition (SER), leading to the adoption of deep architectures for improved performance \cite{mustaqeem2019cnn,kumbhar2019speech,wani2020speech}. Prior to the deep learning era, SER systems primarily relied on classical machine learning methods such as Hidden Markov Models (HMM), Gaussian Mixture Models (GMM), and Support Vector Machines (SVM) \cite{nwe2003speech,patel2017emotion,chen2012speech}. These traditional systems often required extensive preprocessing and manual feature engineering to extract relevant acoustic and prosodic cues.

Feature extraction remains a critical component of SER. Commonly used features include prosodic attributes (e.g., pitch, intensity), voice quality parameters, and spectral descriptors \cite{gangamohan2016analysis}. Among spectral features, Mel-Frequency Cepstral Coefficients (MFCCs) are widely used. For example, \cite{kumbhar2019speech} employed MFCCs with 39 coefficients as input to a Long Short-Term Memory (LSTM) network for emotion classification. Convolutional Neural Networks (CNNs) have also been utilized to extract high-level features from spectrograms \cite{wani2020speech,yenigalla2018speech}. In particular, Deep Stride CNNs (DSCNNs), which replace pooling layers with strided convolutions, have been shown to improve emotion recognition accuracy \cite{wani2020speech,mustaqeem2019cnn}.

Despite these advancements, SER systems continue to struggle with two major challenges: (1) generalization to \textit{unseen cross-corpus data}, and (2) robustness under \textit{realistic noisy conditions} during inference. Many existing approaches attempt to mitigate noise sensitivity using methods such as speech enhancement \cite{zhou20gIS}, noise reduction \cite{Jouni_2016}, feature compensation \cite{chakraborty19IS}, or robust feature extraction techniques \cite{leem2022not,leem2023selective}. However, these approaches often fall short when applied to dynamic and unpredictable acoustic environments.

Similarly, the diversity across emotional speech corpora—including differences in language, culture, recording conditions, and speaker demographics—poses a significant obstacle to cross-corpus generalization. Several techniques have been explored to bridge this gap. These include corpus-based normalization \cite{schuller2010cross}, domain adaptation strategies such as Universum learning \cite{deng2017universum}, and adversarial learning for unsupervised and semi-supervised adaptation \cite{abdelwahab2018domain,latif2020multi}.

While these methods have yielded improvements, they often require additional adaptation stages, access to target domain data, or complex training schemes. This motivates the need for a more streamlined and generalizable approach to SER that is robust against unseen noise and cross-corpus variation without reliance on explicit adaptation.

To this end, our work introduces a two-stage embedding learning strategy: \textit{Emotion-Disentangled Representation Learning} (\edrl) to capture emotion-specific yet generalizable features, followed by \textit{Multiblock Embedding Alignment} \mea\ to refine and align those features in a shared latent space. Together, \edrl\ and \mea\ enable the model to learn robust and transferable representations, improving SER performance in both noisy and cross-corpus settings without requiring target-domain fine-tuning or data augmentation.

\section{Methodology}
\label{sec:method}

\begin{figure}[tb]
	\centering

		\fbox{\includegraphics[width=0.9\textwidth]{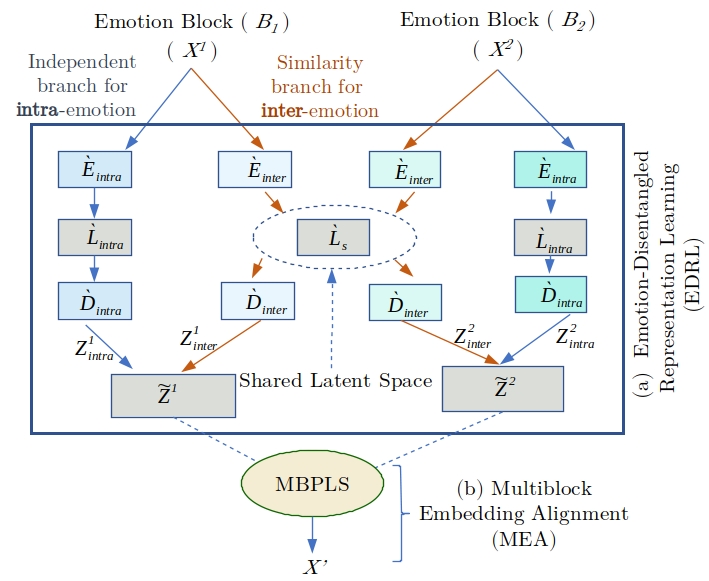}}
		
	\caption{\edrl-\mea\ architecture for $2$ classes.
}
	\label{fig:system_design}
\end{figure}	

\subsection{Emotion-Disentangled Representation Learning (EDRL)}

Emotion-Disentangled Representation Learning (\edrl) 
aims to transform raw speech input \( X \) into a structured embedding space that captures both class-specific emotional traits and shared characteristics across different emotion categories (as depicted in Figure \ref{fig:system_design}(a)). This dual representation facilitates learning of discriminative features while preserving generalizable patterns useful for robust cross-corpus and noisy-condition generalization.

Let:
\begin{itemize}
    \item \( X = \{ X^1, X^2, \dots, X^C \} \) denote the speech input grouped by emotion class \( c \in \{1, \dots, C\} \),
    \item \( C \) be the number of emotion classes, with \( X^c \) containing the samples from class \( c \).
\end{itemize}

For each class \( c \), we define an emotion-specific block \( B_c \) consisting of two parallel encoders:
\begin{itemize}
    \item An \textit{intra-class encoder} (independent branch encoder \( \textit{\`E}_{intra} \) in Figure \ref{fig:system_design}(a)) \( f^{(c)}_{\mathrm{intra}}(\cdot; \theta^{(c)}_{\mathrm{intra}}) \) that learns discriminative features unique to class \( c \),
    \item An \textit{inter-class encoder} (similarity branch encoder \( \textit{\`E}_{inter} \) in Figure \ref{fig:system_design}(a)) \( f^{(c)}_{\mathrm{inter}}(\cdot; \theta^{(c)}_{\mathrm{inter}}, \bar{\theta}_{\mathrm{inter}}) \) that extracts features shared across emotion categories, with \( \bar{\theta}_{\mathrm{inter}} \) being shared across all classes.
\end{itemize}

These encoders produce the following latent representations:
\[
Z_{\mathrm{intra}}^c = f^{(c)}_{\mathrm{intra}}(X^c), \quad
Z_{\mathrm{inter}}^c = f^{(c)}_{\mathrm{inter}}(X^c)
\]

Each block \( B_c \) functions as an autoencoder, where the encoded features are decoded to reconstruct the input. The inter-class latent space is shared, enabling alignment across classes for similarity-aware learning.

Training involves minimizing the reconstruction loss:
\[
\theta^{*(c)}_{\mathrm{intra}}, \theta^{*(c)}_{\mathrm{inter}}, \bar{\theta}^*_{\mathrm{inter}} =
\arg\min \mathcal{L}_r^c \left(
X^c, f^{(c)}_{\mathrm{intra}}(X^c), f^{(c)}_{\mathrm{inter}}(X^c)
\right)
\]

The loss \( \mathcal{L}_r^c \) includes:
\begin{enumerate}
    \item A cosine similarity loss between original and reconstructed embeddings,
    \item A Kullback–Leibler divergence term encouraging compact, disentangled representations.
\end{enumerate}

The joint representation for class \( c \) is obtained by concatenating the intra- and inter-class embeddings:
\[
\widetilde{Z}^c = \left[ Z_{\mathrm{intra}}^c \;\| \; Z_{\mathrm{inter}}^c \right]
\]

These embeddings are passed to the next stage for global alignment.

\subsection{Multiblock Embedding Alignment (MEA)}

Given the combined embeddings \( \{ \widetilde{Z}^c \}_{c=1}^C \), the goal of Multiblock Embedding Alignment (\mea) is to project them into a common latent space that captures both within-class cohesion and between-class similarity structure.

We employ Multiblock Partial Least Squares (MBPLS) to perform this alignment. MBPLS maximizes the covariance between the learned emotion embeddings and the original input features while minimizing redundancy across blocks.

Let:
\begin{itemize}
    \item \( \widetilde{Z} = [ \widetilde{Z}^1, \dots, \widetilde{Z}^C ] \) denote the concatenated embeddings,
    \item \( X \) represent the original speech input,
    \item \( K \) be the number of latent variables (LVs) to be extracted.
\end{itemize}

For each latent variable \( k = 1, \dots, K \), MBPLS computes:
\begin{itemize}
    \item Score vectors \( t_{sk} \) from \( \widetilde{Z}^c \) and \( u_k \) from \( X \),
    \item Loading vectors \( p_k \) and \( v_k \) for the respective components.
\end{itemize}

Embeddings are iteratively updated via deflation:
\[
\widetilde{Z}^{c}_{k+1} = \widetilde{Z}^{c}_k - t_{sk} p_k^\top
\]

After \( K \) iterations, the projected outputs are:
\[
T_s = [t_{s1}, \dots, t_{sK}], \quad
U = [u_1, \dots, u_K], \quad
P = [p_1, \dots, p_K], \quad
V = [v_1, \dots, v_K]
\]

These satisfy the following reconstruction relations:
\[
\widetilde{Z}^c = T_s P_c^\top + E_c, \quad
X = U V^\top + E_X, \quad
X \approx \widetilde{Z} \beta + E
\]

The final \mea\ transformation is defined as:
\[
\phi_{\mathrm{mea}}: \mathrm{MBPLS}([\widetilde{Z}^1, \ldots, \widetilde{Z}^C], X) \rightarrow X'
\]
where \( X' \) denotes the aligned embedding capturing both class structure and its relationship to the original speech signal.

\subsection{Final Classification}

The final representation \( X' \) obtained from the \edrl\ + \mea\ pipeline is fed into a classifier:
\[
\hat{c} = \arg\max_{c} \, P(c \mid X', \Omega)
\]
where \( \hat{c} \) is the predicted emotion class and \( \Omega \) denotes the classifier parameters.

The complete pipeline is summarized in Algorithm~\ref{alg:edrl_mea}.

\begin{algorithm}
\caption{\edrl-\mea: Robust Emotion Representation Learning}
\label{alg:edrl_mea}
\begin{algorithmic}[1]
\Require Speech data $X = \{X^1, X^2, \dots, X^C\}$, where $X^c$ denotes samples from class $c$, with $C$ total classes
\Ensure Robust emotion embeddings $X'$ for classification

\State \textbf{Initialize:} Parameters $\theta^c_{\mathrm{intra}}, \theta^c_{\mathrm{inter}}, \bar{\theta}_{\mathrm{inter}}$ for all $c \in \{1,\dots,C\}$

\ForAll{classes $c = 1$ to $C$}
    \State Extract intra-class embedding: $Z^{c}_{\mathrm{intra}} = f^{(c)}_{\mathrm{intra}}(X^c; \theta^c_{\mathrm{intra}})$
    \State Extract inter-class embedding: $Z^{c}_{\mathrm{inter}} = f^{(c)}_{\mathrm{inter}}(X^c; \theta^c_{\mathrm{inter}}, \bar{\theta}_{\mathrm{inter}})$
    \State Form embedding: $\widetilde{Z}^c = [Z^{c}_{\mathrm{intra}} \;\| \; Z^{c}_{\mathrm{inter}}]$
    \State Minimize reconstruction loss:
    \[
    \theta^{*(c)}_{\mathrm{intra}}, \theta^{*(c)}_{\mathrm{inter}}, \bar{\theta}^*_{\mathrm{inter}} = \arg\min \mathcal{L}_r^c(X^c, Z^{c}_{\mathrm{intra}}, Z^{c}_{\mathrm{inter}})
    \]
\EndFor

\State \textbf{Input to \mea:} Embeddings $\widetilde{Z} = \{\widetilde{Z}^1, \dots, \widetilde{Z}^C\}$ and original input $X$
\State Apply MBPLS projection: $X' = \phi_{\mathrm{mea}}(\widetilde{Z}, X)$
\State \textbf{Emotion Classification:} $\hat{c} = \arg\max_c P(c \mid X', \Omega)$
\end{algorithmic}
\end{algorithm}

\section{Experimental Setup}
\label{sec:exp}
\label{exp}
We performed the evaluation of our proposed approach using intra-corpus as well as inter-corpus setup in clean and noisy environments, respectively. In real life conversations, e.g. in call center help-desk or mental-health screening, emotions are mostly interpreted as positive or negative in dimensional space. That is why in this paper we choose to experimentally validate our proposed approach in arousal and valence dimensions.   
 \subsection{Database and \Feature\ Extraction}
\label{sec:data}

\noindent {\bf SER Database:} We use Interactive emotional dyadic motion capture (IEMOCAP) database \cite{Iemocap}, wherein samples were recorded when two participants conversing in two different scenarios, namely scripted and improvised. In scripted sessions, the speakers were asked to memorize the scripts and rehearse, whereas in improvised they were asked to improvise some hypothetical situations that were designed to elicit the specific emotions. 
Each samples are annotated by participants themselves and by several evaluators in 10 emotion categories as well as in \Aro-\Val-\Dom\ dimensional space on a scale of 1 to 5.
In this work, we perform the binary-class SER in dimensional emotion space. 
We consider the average of all evaluators score as a final rating given to each sample.
Furthermore, we construct the binary labels \high\ ($\ge \lambda$) and \low\ ($<\lambda$) on the final rating score with  $\lambda = 2.5$.
Thus 
each sample, had one of two labels in the  \Val-\Aro-\ space, namely, \Val\high\  or \Val\low; \Aro\high\ or \Aro\low.

For inter-corpus evaluation, we combined two audio emotion datasets, namely, (A) Berlin Emotional Database ({\sc Emo-DB}) \cite{EmoDB}
and, (B) the Ryerson Audio-Visual Database of Emotional Speech and Song ({\sc RAVDESS}) \cite{livingstone2018ryerson}. 
To match with our train setup, while performing the inter-corpus evaluation in dimensional space, we consider four categorical emotion classes {\tt Anger, \tt Happy, \tt Neutral and \tt Sad} that are mapped into \Aro-\Val\ space.
This is done by labeling {\tt Anger and \tt Sad} as \Val\low, {\tt Happy and \tt Neutral} as \Val\high, {\tt Sad and \tt Neutral} as \Aro\low\ and {\tt Anger and \tt Happy} as \Aro\high. This resulted into data distribution of \Val\high: 438, \Val\low: 573
\Aro\high: 582, \Aro\low: 429.

\noindent {\bf Noise Database:} In order to create a noisy test data, 
we use recorded noises from Indian Noise Database (iNoise) database \cite{iNoise} to corrupt the
clean test utterances for both inter-corpus and intra-corpus setup. We used total five types of noises, out of which 3 noises are indoor, namely, {\tt Indoor\_workplace}, {\tt Indoor\_cafteria}, {\tt Indoor\_home}, represented as \N1, \N2\ and \N3, respectively; and 2 outdoor noises, namely, {\tt Outdoor\_travel-bus}, {\tt Outdoor\_street}, represented as \N4\ and \N5, respectively. All these noises are used to corrupt clean test samples at 5
SNR levels (0dB, 5dB, 10dB, 15dB, 20dB). This is to be noted that the choice of noise types are made in such a way that the environments are closely relevant to real life scenarios.

\noindent {\bf Acoustic \Feature\ Extraction:} We extract 
$88$ 
acoustic features 
from each audio file with extended Geneva Minimalistic Acoustic Parameter Set (eGeMAPS)  using eGeMAPSv01a \cite{eyben2015geneva}
configuration file of 
the
OpenSMILE toolkit \cite{opensmile}. 
There are a total of $18$ acoustic features, namely
  Pitch, Jitter, Shimmer, formant related energy, MFCCs,  
  also known as  low-level descriptors (LLDs);
and  high-level descriptors (HLDs) are computed (mean, standard deviation, skewness, kurtosis, extremes, linear regressions, etc.) for each of those LLDs. 

\subsection{EDRL-MEA Configuration and Training}
\label{models}
We implement \edrl\ with two emotion blocks ($B_1$ and $B_2$) as shown in Figure~\ref{fig:system_design}(b). \AE\ %
for both \textit{intra} and \textit{inter} branch %
consists of $3$ layers with {\em relu} activation, namely, \E, \L\ and \D.
To keep the \AE\ compact, we stacked only single \E, \L\ and \D\ layers in
each branch. We tried different setups for selecting the number of hidden neurons in each layer;
(a) setup-1: same number of neurons in \E, \L\ and \D; (b) setup-2: compressed latent space with
number of hidden neurons in \L\  as half of that in \E\ and \D; and (c) setup-3: expanded latent space with
number of hidden neurons in \L\ as twice of that in \E\ and \D . Each of the above mentioned setups are
tried with $N/2$, $N$, $2N$, and $4N$ number of hidden neurons, where $N$ is dimension of the input vector. We found setup-3 with $2N$ neurons to be working best.
Further, the decoded output from both the branches are concatenated and fed to the final dense layer with {\em linear} activation and $N$ neurons. The final output of each \edrl\ block is the combined representation learnt per emotion class. So, we get %
 one $N$-dimension \edrl\ output vectors %
 for each block. \L\ of the similarity branches from two blocks are tied together (by sharing weights) unlike the independent branches of the two block.
 {\em We hypothesize that this process of learning the two branches helps the model capture not only the emotion 
 class specific properties but also similarities among %
 the different emotion classes.}
As an example, assume $X_{\train}$ be the training set that consists of two class data $X_{\train}^1$ and $X_{\train}^2$. During training, $X_{\train}^1$ is input to block $B_1$ (in an epoch) while $X_{\train}^2$ is input to $B_2$ in a sequence. %
 At each epoch $e$, both $B_1$ and $B_2$ are trained. While the shared latent space weights are updated 
 by training each block for an epoch, the  %
 block layer weights are updated only once per epoch when that block sees an input. The \edrl\ is %
 implemented in {\em Keras} \cite{Keras} with {\em adam} optimizer and customised loss. 
 To prevent overfitting, we use Keras \textit{EarlyStopping} that monitors validation loss to guide \edrl\ training.
We use  a python package, \textit{mbpls}, to implement the \mea\ with two data blocks consisting of $N$-dimensional combined embeddings learnt from both $B_1$ and $B_2$ of \edrl. Note that \mea\ is trained on \edrl\ output %
 and maps them to a common latent subspace. The target vector of \mea\ is the original train data itself. 
From these two data blocks, \mea\ predicts a $N$-dimensional vector, such that respective contribution of each emotion block is retained. Finally, this emotion class embedding is used for emotion classification.

\subsection{Emotion Classification}
\label{sec:results}
We split the IEMOCAP data into train set ($80\%$) and  test set ($20\%$) for training and intra-corpus testing, respectively. 
Further $10\%$ of the train data is used for validation for \edrl-\mea\ training.
IEMOCAP dataset consists of \Val+: 2952, \Val-: 2483; \Aro+: 3480, \Aro-: 1995 samples. We adopted majority class undersampling using {\tt RandomUnderSampler} technique (from sklearn python package) over the train set to overcome the class imbalance across the \Val-\Aro\
emotion dimensions. Unlike conventional approach of data balancing which uses minority class oversampling, we opted for majority class undersampling to avoid synthetically generated samples to be used in training.
We build two SER systems, (1) Baseline SER system using the features mentioned in Section \ref{sec:data} and Random Forest (\RF) as the final stage classifier; (2) \edrl-\mea\ based SER system that uses the 
reconstructed embedding $X'$ ( as represented in Figure \ref{fig:system_design}) learnt using the proposed approach to perform the classification using \RF.
We use \RF\ for the final emotion recognition task because of it's superior performance compared to other standard classifiers like \SVM, \KNN, and \ANN.

\subsection{Experimental Results and Analysis}

\begin{table}[!t]
\centering
\caption{Intra-corpus Performance (F1 score) of Clean model (Baseline vs \edrl\-\mea) with clean and noisy test data; 
(Train and Test dataset both are from IEMOCAP)}
\label{tab:edrl-mea_intra_noisy}
\resizebox{\columnwidth}{!}{
\begin{tabular}{|cc|ll|ll|}
\hline
\multicolumn{1}{|c|}{\multirow{2}{*}{\textbf{Environment}}} & \multirow{2}{*}{\textbf{Noise\_type}} & \multicolumn{2}{c|}{\textbf{\Aro}} & \multicolumn{2}{c|}{\textbf{\Val}} \\ \cline{3-6} 
\multicolumn{1}{|c|}{} &  & \multicolumn{1}{c|}{Baseline} & \multicolumn{1}{c|}{\edrl-\mea} & \multicolumn{1}{c|}{Baseline} & \multicolumn{1}{c|}{\edrl-\mea} \\ \hline
\multicolumn{2}{|c|}{\textbf{Clean}} & \multicolumn{1}{l|}{77.7} & 80.1 (+2.4) & \multicolumn{1}{l|}{66.7} & 70.6 (+3.9) \\ \hline
\multicolumn{1}{|c|}{\multirow{5}{*}{\textbf{Noisy}}} & \N1\ & \multicolumn{1}{l|}{52.72} & 54.64 (+1.92) & \multicolumn{1}{l|}{47.06} & 50.72 (+3.66) \\ \cline{2-6} 
\multicolumn{1}{|c|}{} & \N2\ & \multicolumn{1}{l|}{52.26} & 53.74 (+1.48) & \multicolumn{1}{l|}{47.46} & 49.4 (+1.94) \\ \cline{2-6} 
\multicolumn{1}{|c|}{} & \N3\ & \multicolumn{1}{l|}{52.78} & 54.86 (+2.08) & \multicolumn{1}{l|}{46.02} & 49.88 (+3.86) \\ \cline{2-6} 
\multicolumn{1}{|c|}{} & \N4\ & \multicolumn{1}{l|}{51.6} & 54.26 (+2.66) & \multicolumn{1}{l|}{46.6} & 49.04 (+2.44) \\ \cline{2-6} 
\multicolumn{1}{|c|}{} & \N5\ & \multicolumn{1}{l|}{50.82} & 54.7 (+3.88) & \multicolumn{1}{l|}{47.46} & 50.34 (+2.88) \\ \hline
\end{tabular}}
\end{table}
\begin{table}[!t]
\centering
\caption{Inter-corpus Performance (F1 score) of Clean model (Baseline vs \edrl\-\mea) with clean and noisy test data; 
(Train dataset: IEMOCAP; Test dataset: EMODB+RAVDESS)}
\label{tab:edrl-mea_inter_noisy}
\resizebox{\columnwidth}{!}{
\begin{tabular}{|cc|ll|ll|}
\hline
\multicolumn{1}{|c|}{\multirow{2}{*}{\textbf{Environment}}} & \multirow{2}{*}{\textbf{Noise\_type}} & \multicolumn{2}{c|}{\textbf{\Aro}} & \multicolumn{2}{c|}{\textbf{\Val}} \\ \cline{3-6} 
\multicolumn{1}{|c|}{} &  & \multicolumn{1}{c|}{Baseline} & \multicolumn{1}{c|}{\edrl-\mea} & \multicolumn{1}{c|}{Baseline} & \multicolumn{1}{c|}{\edrl-\mea} \\ \hline
\multicolumn{2}{|c|}{\textbf{Clean}} & \multicolumn{1}{l|}{56.8} & 65.3 (+8.5) & \multicolumn{1}{l|}{54.1} & 60.2 (+6.1) \\ \hline
\multicolumn{1}{|c|}{\multirow{5}{*}{\textbf{Noisy}}} & \N1\ & \multicolumn{1}{l|}{54.26} & 56.52 (+2.26) & \multicolumn{1}{l|}{55.14} & 58 (+2.86) \\ \cline{2-6} 
\multicolumn{1}{|c|}{} & \N2\ & \multicolumn{1}{l|}{52.2} & 56.4 (+4.2) & \multicolumn{1}{l|}{50.76} & 56.66 (+5.9) \\ \cline{2-6} 
\multicolumn{1}{|c|}{} & \N3\ & \multicolumn{1}{l|}{50.32} & 55.26 (+4.94) & \multicolumn{1}{l|}{53.2} & 56.96 (+3.76) \\ \cline{2-6} 
\multicolumn{1}{|c|}{} & \N4\ & \multicolumn{1}{l|}{45.52} & 47.74 (+2.22) & \multicolumn{1}{l|}{50.64} & 52.6 (+1.96) \\ \cline{2-6} 
\multicolumn{1}{|c|}{} & \N5\ & \multicolumn{1}{l|}{42.8} & 46.34 (+3.54) & \multicolumn{1}{l|}{49.2} & 54.86 (+5.66) \\ \hline
\end{tabular}}
\end{table}

We evaluate the proposed \edrl-\mea\ approach for SER using intra- and inter-corpus test data from clean as well as noisy environments separately, for both \Val\ and \Aro\ dimensions (as shown in Table \ref{tab:edrl-mea_intra_noisy}, \ref{tab:edrl-mea_inter_noisy}). We use 
\RF\ as the  final stage classifier 
in 
all our experiments. 
We perform grid search to fix the \RF\ parameters \textit{n\_estimators} and \textit{n\_depth} for each of our experimental setup independently, with grid of $n\_estimators= (i*10)$, where $50 \le i \le 500$, and $n\_depth= (2*i)$, where $1 \le i \le 20$. 
It is to be noted that both Baseline and \edrl-\mea\ system are trained using
IEMOCAP clean samples from the training set. Furthermore, the trained clean model (for Baseline vs \edrl-\mea) is evaluated in four different setup. We discuss each experimental setup in brief details as below.

\begin{enumerate}
    \item{Intra-corpus Clean:} Both Baseline and \edrl-\mea\ is tested using clean test set from IEMOCAP. As shown in Table \ref{tab:edrl-mea_intra_noisy}, \edrl-\mea\ surpasses the Baseline in terms of F1 score, with absolute improvement of 2.4\% and 3.9\% for \Aro\ and \Val, respectively.
    
    \item{Intra-corpus Noisy:} Firstly, noisy test data is prepared by corrupting IEMOCAP test set using 5 noise-types from iNoise dataset at 5 SNR levels (as discussed in Section \ref{sec:data}). We report average F1-score over 5 SNR levels for each noise-type as seen in Table \ref{tab:edrl-mea_intra_noisy}. \edrl-\mea\ shows an absolute improvement over Baseline for both \Aro-\Val\ emotions in terms of
    F1 scores across all 5 noise-types. There is an absolute improvement of (\Aro:1.92\%, \Val:3.66\%), (\Aro:1.48\%, \Val:1.94\%), (\Aro:2.08\%, \Val:3.86\%), (\Aro:2.66\%, \Val:2.44\%) and (\Aro:3.88\%, \Val:2.88\%) using noisy test set corrupted with \N1, \N2, \N3, \N4\ and \N5, respectively.

    \item{Inter-corpus Clean:} This setup is used to test the cross-corpus generalization of \edrl-\mea\ embeddings. As discussed in Section \ref{sec:data}, EmoDB and RAVDESS dataset are combined together to form an inter-corpus test set. Our proposed approach outperforms the Baseline even in cross-corpus testing with a significant improvement of 8.5\% and 6.1\% in terms of F1 score for \Aro\ and \Val\ emotion, respectively, as seen in Table \ref{tab:edrl-mea_inter_noisy}.

    \item{Inter-corpus Noisy:} Similar to intra-corpus noisy data, inter-corpus noisy data is prepared by corrupting the inter-corpus clean test set with 
    same noises and SNR levels. For each noise type, we report the performance as an average F1 score over 5 SNR level. As shown in Table \ref{tab:edrl-mea_inter_noisy}, \edrl-\mea\ surpasses the Baseline with an absolute improvement of (\Aro:2.26\%, \Val:2.86\%), (\Aro:4.2\%, \Val:5.9\%), (\Aro:4.94\%, \Val:3.76\%), (\Aro:2.22\%, \Val:1.96\%) and (\Aro:3.54\%, \Val:5.66\%) using noisy test set corrupted with \N1, \N2, \N3, \N4\ and \N5, respectively.
\end{enumerate}

The SER performance using \edrl-\mea\ not only surpasses the
Baseline in clean intra-corpus setup, but also shows a significant improvement over Baseline in noisy environment and cross-corpus testing, clearly demonstrates the usefulness of the proposed approach.
It is to be noted that in this paper we are not aiming for any multi-conditioning based model adaptation to address the noise aspect in the speech data. We show the effectiveness of the learnt embeddings with proposed \edrl-\mea\ in both cross-corpus and noisy environment settings, restricting the model training only with the clean data. As can be seen, the resultant embeddings capture intra- and inter-class characteristics, benefit the final-stage classifier with an improved SER performance, through better generalization  on cross-corpus data and more robustness to the unseen noises.
Please note that we make no effort to compare our
results with existing work \cite{wagner2023dawn,lu2018learning,li2021contrastive} on SER 
in dimensional emotion space, due to the mismatch in experimental setup as compared to ours.

\section{Conclusion}
\label{sec:conclu}
This paper introduces an effective two-stage framework for robust speech emotion recognition (SER) under cross-corpus and noisy conditions. Our approach integrates Emotion-Disentangled Representation Learning (\edrl) to simultaneously capture emotion-specific and shared inter-class patterns through parallel intra- and inter-class encoding pathways. This disentanglement encourages the model to learn discriminative yet generalizable embeddings that are less sensitive to corpus-specific or noise-related artifacts. To further enhance robustness, we incorporate Multiblock Embedding Alignment (\mea) using Multiblock Partial Least Squares (MBPLS), which aligns the learned embeddings with the original input space. This projection mechanism preserves both intra-class distinctiveness and inter-class consistency, ensuring that the embeddings remain semantically meaningful even under distributional shifts. Experimental results validate that the proposed \edrl+\mea\ pipeline significantly outperforms competitive baselines in both cross-corpus and noisy evaluation setups. These findings demonstrate the effectiveness of our method in mitigating the adverse effects of unseen noise and corpus variability, a critical requirement for real-world SER systems. Our work contributes a generalizable and noise-resilient modeling paradigm, paving the way for more reliable affective computing applications in diverse and unconstrained environments.

 \bibliographystyle{splncs04}
 \bibliography{mybibliography}

\begin{thebibliography}{10}
\providecommand{\url}[1]{\texttt{#1}}
\providecommand{\urlprefix}{URL }
\providecommand{\doi}[1]{https://doi.org/#1}

\bibitem{opensmile}
open{SMILE}, audio feature extraction tool by aud{EERING}. http://www.audeering.com/ research/opensmile, accessed: 2019

\bibitem{abdelwahab2018domain}
Abdelwahab, M., Busso, C.: Domain adversarial for acoustic emotion recognition. IEEE/ACM Transactions on Audio, Speech, and Language Processing  \textbf{26}(12),  2423--2435 (2018)

\bibitem{abhishek2023wecareimprovingcode}
Abhishek, N.V.S., Bhattacharyya, P.: "we care": Improving code mixed speech emotion recognition in customer-care conversations (2023), \url{https://arxiv.org/abs/2308.03150}

\bibitem{ahn2021cross}
Ahn, Y., Lee, S.J., Shin, J.W.: Cross-corpus speech emotion recognition based on few-shot learning and domain adaptation. IEEE Signal Processing Letters  \textbf{28},  1190--1194 (2021)

\bibitem{braunschweiler2021study}
Braunschweiler, N., Doddipatla, R., Keizer, S., Stoyanchev, S.: A study on cross-corpus speech emotion recognition and data augmentation. In: 2021 IEEE Automatic Speech Recognition and Understanding Workshop (ASRU). pp. 24--30. IEEE (2021)

\bibitem{EmoDB}
Burkhardt, F., Paeschke, A., Rolfes, M.A., Sendlmeier, W.F., Weiss, B.: A {D}atabase of {G}erman {E}motional {S}peech. In: Proc. Interspeech (2005)

\bibitem{Iemocap}
Busso, C., Bulut, M., Lee, C.C., Kazemzadeh, A., Mower, E., Kim, S., Chang, J.N., Lee, S., Narayanan, S.S.: {IEMOCAP}: Interactive emotional dyadic motion capture database. Language resources and evaluation  \textbf{42}(4), ~335 (2008)

\bibitem{chakraborty19IS}
Chakraborty, R., Panda, A., Pandharipande, M., Joshi, S., Kopparapu, S.K.: Front-end feature compensation and denoising for noise robust speech emotion recognition. In: Interspeech 2019. pp. 3257--3261 (2019). \doi{10.21437/Interspeech.2019-2243}

\bibitem{chen2012speech}
Chen, L., Mao, X., Xue, Y., Cheng, L.L.: Speech emotion recognition: Features and classification models. Digital signal processing  \textbf{22}(6),  1154--1160 (2012)

\bibitem{deng2017universum}
Deng, J., Xu, X., Zhang, Z., Fr{\"u}hholz, S., Schuller, B.: Universum autoencoder-based domain adaptation for speech emotion recognition. IEEE Signal Processing Letters  \textbf{24}(4),  500--504 (2017)

\bibitem{eyben2015geneva}
Eyben, F., Scherer, K.R., Schuller, B.W., Sundberg, J., Andr{\'e}, E., Busso, C., Devillers, L.Y., Epps, J., Laukka, P., Narayanan, S.S., et~al.: The geneva minimalistic acoustic parameter set (gemaps) for voice research and affective computing. IEEE transactions on affective computing  \textbf{7}(2),  190--202 (2015)

\bibitem{fahad2021survey}
Fahad, M.S., Ranjan, A., Yadav, J., Deepak, A.: A survey of speech emotion recognition in natural environment. Digital signal processing  \textbf{110},  102951 (2021)

\bibitem{gangamohan2016analysis}
Gangamohan, P., Kadiri, S.R., Yegnanarayana, B.: Analysis of emotional speech—a review. Toward Robotic Socially Believable Behaving Systems-Volume I: Modeling Emotions pp. 205--238 (2016)

\bibitem{george2024review}
George, S.M., Ilyas, P.M.: A review on speech emotion recognition: a survey, recent advances, challenges, and the influence of noise. Neurocomputing  \textbf{568},  127015 (2024)

\bibitem{Keras}
Keras: K{ERAS}: The python deep learning library. https://keras.io/ (2019), accessed: 2019

\bibitem{iNoise}
Kopparapu, S.K., Sheikh, I., Thanneeru, V.K.: inoise indian noise database (2020). \doi{10.21227/w3xm-jn45}, \url{https://dx.doi.org/10.21227/w3xm-jn45}

\bibitem{kumbhar2019speech}
Kumbhar, H.S., Bhandari, S.U.: Speech emotion recognition using mfcc features and lstm network. In: 2019 5th international conference on computing, communication, control and automation (ICCUBEA). pp.~1--3. IEEE (2019)

\bibitem{latif2020multi}
Latif, S., Rana, R., Khalifa, S., Jurdak, R., Epps, J., Schuller, B.W.: Multi-task semi-supervised adversarial autoencoding for speech emotion recognition. IEEE Transactions on Affective computing  \textbf{13}(2),  992--1004 (2020)

\bibitem{leem2022not}
Leem, S.G., Fulford, D., Onnela, J.P., Gard, D., Busso, C.: Not all features are equal: Selection of robust features for speech emotion recognition in noisy environments. In: ICASSP 2022-2022 IEEE International Conference on Acoustics, Speech and Signal Processing (ICASSP). pp. 6447--6451. IEEE (2022)

\bibitem{leem2023selective}
Leem, S.G., Fulford, D., Onnela, J.P., Gard, D., Busso, C.: Selective acoustic feature enhancement for speech emotion recognition with noisy speech. IEEE/ACM Transactions on Audio, Speech, and Language Processing  (2023)

\bibitem{li2021contrastive}
Li, M., Yang, B., Levy, J., Stolcke, A., Rozgic, V., Matsoukas, S., Papayiannis, C., Bone, D., Wang, C.: Contrastive unsupervised learning for speech emotion recognition. In: ICASSP 2021-2021 IEEE International Conference on Acoustics, Speech and Signal Processing (ICASSP). pp. 6329--6333. IEEE (2021)

\bibitem{lin2024selfcontextawareemotionperception}
Lin, Z., Cruz, F., Sandoval, E.B.: Self context-aware emotion perception on human-robot interaction (2024), \url{https://arxiv.org/abs/2401.10946}

\bibitem{livingstone2018ryerson}
Livingstone, S.R., Russo, F.A.: The ryerson audio-visual database of emotional speech and song (ravdess): A dynamic, multimodal set of facial and vocal expressions in north american english. PloS one  \textbf{13}(5),  e0196391 (2018)

\bibitem{lu2018learning}
Lu, C.C., Li, J.L., Lee, C.C.: Learning an arousal-valence speech front-end network using media data in-the-wild for emotion recognition. In: Proceedings of the 2018 on Audio/Visual Emotion Challenge and Workshop. pp. 99--105 (2018)

\bibitem{Mamun_2023}
Mamun, M.A., Abdullah, H.M., Alam, M.G.R., Hassan, M.M., Uddin, M.Z.: Affective social anthropomorphic intelligent system. Multimedia Tools and Applications  \textbf{82}(23),  35059–35090 (Mar 2023). \doi{10.1007/s11042-023-14597-6}, \url{http://dx.doi.org/10.1007/s11042-023-14597-6}

\bibitem{mustaqeem2019cnn}
Mustaqeem, Kwon, S.: A cnn-assisted enhanced audio signal processing for speech emotion recognition. Sensors  \textbf{20}(1), ~183 (2019)

\bibitem{nwe2003speech}
Nwe, T.L., Foo, S.W., De~Silva, L.C.: Speech emotion recognition using hidden markov models. Speech communication  \textbf{41}(4),  603--623 (2003)

\bibitem{parry2019analysis}
Parry, J., Palaz, D., Clarke, G., Lecomte, P., Mead, R., Berger, M., Hofer, G.: Analysis of deep learning architectures for cross-corpus speech emotion recognition. In: Interspeech. pp. 1656--1660 (2019)

\bibitem{patel2017emotion}
Patel, P., Chaudhari, A., Kale, R., Pund, M.: Emotion recognition from speech with gaussian mixture models \& via boosted gmm. International Journal of Research In Science \& Engineering  \textbf{3} (2017)

\bibitem{Jouni_2016}
Pohjalainen, J., Fabien~Ringeval, F., Zhang, Z., Schuller, B.: Spectral and cepstral audio noise reduction techniques in speech emotion recognition. In: Proceedings of the 24th ACM International Conference on Multimedia. p. 670–674. MM '16, Association for Computing Machinery, New York, NY, USA (2016). \doi{10.1145/2964284.2967306}, \url{https://doi.org/10.1145/2964284.2967306}

\bibitem{rathi2024analyzing}
Rathi, T., Tripathy, M.: Analyzing the influence of different speech data corpora and speech features on speech emotion recognition: A review. Speech Communication p. 103102 (2024)

\bibitem{schuller2010cross}
Schuller, B., Vlasenko, B., Eyben, F., W{\"o}llmer, M., Stuhlsatz, A., Wendemuth, A., Rigoll, G.: Cross-corpus acoustic emotion recognition: Variances and strategies. IEEE Transactions on Affective Computing  \textbf{1}(2),  119--131 (2010)

\bibitem{Schuller-STOA2018}
Schuller, B.W.: Speech emotion recognition: two decades in a nutshell, benchmarks, and ongoing trends. Commun. ACM  \textbf{61}(5),  90–99 (Apr 2018). \doi{10.1145/3129340}, \url{https://doi.org/10.1145/3129340}

\bibitem{TiwariCK24}
Tiwari, U., Chakraborty, R., Kopparapu, S.K.: Joint class learning with self similarity projection for {EEG} emotion recognition. In: Natarajan, S., Bhattacharya, I., Singh, R., Kumar, A., Ranu, S., Bali, K., K, A. (eds.) Proceedings of the 7th Joint International Conference on Data Science {\&} Management of Data (11th {ACM} {IKDD} {CODS} and 29th COMAD), Bangalore, India, January 4-7, 2024. pp. 207--211. {ACM} (2024). \doi{10.1145/3632410.3632417}, \url{https://doi.org/10.1145/3632410.3632417}

\bibitem{wagner2023dawn}
Wagner, J., Triantafyllopoulos, A., Wierstorf, H., Schmitt, M., Burkhardt, F., Eyben, F., Schuller, B.W.: Dawn of the transformer era in speech emotion recognition: closing the valence gap. IEEE Transactions on Pattern Analysis and Machine Intelligence  \textbf{45}(9),  10745--10759 (2023)

\bibitem{wani2020speech}
Wani, T.M., Gunawan, T.S., Qadri, S.A.A., Mansor, H., Kartiwi, M., Ismail, N.: Speech emotion recognition using convolution neural networks and deep stride convolutional neural networks. In: 2020 6th International Conference on Wireless and Telematics (ICWT). pp.~1--6. IEEE (2020)

\bibitem{yenigalla2018speech}
Yenigalla, P., Kumar, A., Tripathi, S., Singh, C., Kar, S., Vepa, J.: Speech emotion recognition using spectrogram \& phoneme embedding. In: Interspeech. vol.~2018, pp. 3688--3692 (2018)

\bibitem{zhou20gIS}
Zhou, H., Du, J., Tu, Y.H., Lee, C.H.: Using speech enhancement preprocessing for speech emotion recognition in realistic noisy conditions. In: Interspeech 2020. pp. 4098--4102 (2020). \doi{10.21437/Interspeech.2020-2472}

\end{thebibliography}

\end{document}